\documentclass[twocolumn,prl,tightenlines,superscriptaddress]{revtex4-1}

\usepackage{hyperref}
\usepackage{amsmath,amssymb}
\usepackage[utf8]{inputenc} 			% Encodage UTF-8
\usepackage[T1]{fontenc}				% Encodage du fichier de sortie
\usepackage{graphicx}
\usepackage{float}
\usepackage[usenames,dvipsnames]{xcolor}
\usepackage{tikz}

\newcommand{\eps}{\varepsilon}

\renewcommand{\vec}[1]{\mathbf{#1}}

\newcommand*\colvec[3][]{
    \begin{pmatrix}\ifx\relax#1\relax\else#1\\\fi#2\\#3\end{pmatrix}
}

\begin{document}
%\title{Active Clock Models: Relevance of spatial anisotropy for polar flocks}
%\title{Active Clock Models : Susceptibility of Polar Flocks to Spatial Anisotropy}
\title{Susceptibility of Polar Flocks to Spatial Anisotropy}

\author{Alexandre Solon}
\affiliation{Sorbonne Universit\'e, CNRS, Laboratoire de Physique Th\'eorique de la Mati\`ere Condens\'ee, 75005 Paris, France}

\author{Hugues Chat\'{e}}
\affiliation{Service de Physique de l'Etat Condens\'e, CEA, CNRS Universit\'e Paris-Saclay, CEA-Saclay, 91191 Gif-sur-Yvette, France}
\affiliation{Computational Science Research Center, Beijing 100094, China}
\affiliation{Sorbonne Universit\'e, CNRS, Laboratoire de Physique Th\'eorique de la Mati\`ere Condens\'ee, 75005 Paris, France}

\author{John Toner}
\affiliation{Department of Physics and Institute for Fundamental Science, University of Oregon, Eugene, OR 97403}

\author{Julien Tailleur}
\affiliation{Université de Paris, Laboratoire Matière et Systèmes Complexes (MSC), UMR 7057 CNRS, 75205 Paris, France}

\date{\today}
\begin{abstract}
  We consider the effect of spatial anisotropy on polar flocks by
  investigating active $q$-state clock models in two dimensions. In
  contrast to what happens in equilibrium, we find that, in the
  large-size limit, any amount of anisotropy changes drastically the
  phenomenology of the rotationally-invariant case, destroying
  long-range correlations, pinning the direction of global order, and
  transforming the traveling bands of the coexistence phase into a
  single moving domain. All this happens beyond a lengthscale that
  diverges in the $q\to\infty$ limit. A phenomenology akin to that
  of the Vicsek model can thus be observed in a finite system for
  large enough values of $q$. We provide a scaling argument
  which rationalizes why anisotropy has so different effects in the
  passive and active cases.
  \end{abstract}

\maketitle

Active matter, being made of energy-consuming units, is well known to
exhibit spectacular collective behaviors not permitted in
equilibrium. Experimental examples include the complex defect dynamics of active
nematics~\cite{sanchez_spontaneous_2012,li_data-driven_2019,duclos_topological_2020}, 
low Reynolds number
turbulence~\cite{wensink_meso-scale_2012,martinez_combined_2020}, 
motility-induced phase
separation~\cite{buttinoni_dynamical_2013,geyer_freezing_2019,van_der_linden_interrupted_2019}
and, perhaps most famously,
flocking~\cite{cavagna_scale-free_2010,bricard_emergence_2013,geyer_sounds_2018,deseigne_collective_2010,kumar_flocking_2014}. Although
these phenomena appear in complex, usually living, systems, most of our
theoretical understanding comes from studying collections of identical
active units evolving in pristine environments, often with periodic
boundary conditions. Recently acquired evidence suggests, though, that
active systems seem to be fundamentally sensitive to quenched and
population disorder~\cite{chepizhko_optimal_2013,toner_swarming_2018,toner_hydrodynamic_2018,
  dor_ramifications_2019,duan_breakdown_2021,ventejou_susceptibility_2021,ro_disorder-induced_2021},
and that even the nature of boundaries can influence bulk properties
\cite{dor_far-reaching_2021}.

The sensitivity of active systems to anisotropy, in the form
of fixed preferred directions in space, remains largely unexplored. A
basic result is available in the context of polar flocks, {\it i.e.}
collections of simple self-propelled particles locally aligning their
velocities. In two space dimensions, comparing the Vicsek model
(VM)~\cite{vicsek_novel_1995} to the active Ising model
(AIM)~\cite{solon_revisiting_2013} shows that the symmetry of the
order parameter controls the emerging physics. In the VM, the self
propulsion dynamics are rotation invariant, i.e., have continuous
symmetry, and the ordered phase exhibits scale-free density and order
fluctuations~\cite{toner_long-range_1995, toner_flocks_1998,
  toner_hydrodynamics_2005, toner_reanalysis_2012,
  mahault_quantitative_2019}. In the AIM, directed motion happens only
along two opposite directions, hence the dynamics only has a discrete
symmetry, and the correlations are short ranged in the ordered phase.
Concomitantly, even though the transition to collective motion is akin
to a phase-separation scenario in both the AIM and the VM, their
coexistence phases are different~\cite{solon_phase_2015}: models in
the Vicsek class exhibit microphase separation, typically in the form
of a smectic train of traveling dense
bands~\cite{chate_collective_2008}, whereas the AIM shows a single
moving domain and macrophase separation~\cite{solon_flocking_2015}.

% \if{For
% the Vicsek model (VM)~\cite{vicsek_novel_1995}, \red{with its rotation-invariant dynamics} \red{ \%} %where theself-propulsion velocities satisfy an $SO(2)$ symmetry, 
% one observes
% long-range correlated density and order \red{\parameter} fluctuations in a homogeneous
% flock with true long-range order~\cite{mahault_quantitative_2019},
% something that is prohibited to aligning passive
% particles~\cite{toner_long-range_1995,toner_reanalysis_2012,chate_dry_2020}. For
% the active Ising model (AIM)~\cite{solon_revisiting_2013}, where
% directed motion happens only along two opposite directions and admits
% a $Z_2$ symmetry, the correlations are short ranged in the ordered
% phase. Concomitantly, even though the transition to collective motion
% is akin to a phase-separation scenario in both the AIM and the VM,
% their coexistence phases are different~\cite{solon_phase_2015}: models
% in the Vicsek class exhibit microphase separation, typically in the
% form of a smectic train of traveling dense
% bands~\cite{chate_collective_2008}, whereas the AIM shows a single
% moving domain (macrophase separation)~\cite{solon_flocking_2015}.
% }\fi

The AIM, by restricting directed motion along one dimension,
corresponds to an extreme spatial anisotropy.  A natural question is
then whether polar flocks are affected by weaker forms of anisotropy.
In equilibrium, the phenomenology of the 2D XY model---which is the
passive counterpart of the VM---is in a sense both robust and
sensitive to the discreteness of spins: $q$-state clock models, which
break rotational invariance and interpolate between the XY and the
Ising models, exhibit a quasi-long-range ordered phase similar to that
of the XY model below the BKT transition for $q>4$, but this critical
phase gives way to a region of long-range order below some finite
temperature that vanishes only when
$q\to\infty$~\cite{jose_renormalization_1977,elitzur_phase_1979,tobochnik_properties_1982,lapilli_universality_2006,li_critical_2020}.
Thus, from the XY viewpoint, a new ordered phase emerges at any $q$, but is
marginal, confined to $T=0$, in the $q\to\infty$ limit.

In this Letter, we investigate the susceptibility of polar flocks to
weak anisotropy.  Using a combination of numerical simulations and
analytical arguments, we study $q$-state active clock models and their
hydrodynamic theories.  We uncover a scenario qualitatively different
from the equilibrium one: the phenomenology of the rotationally
invariant Vicsek model disappears for any amount of spin anisotropy,
leaving only AIM-like phenomenology with short-ranged correlations and
macrophase separation.  This, however, happens only asymptotically: at
fixed $q$, one still observes the Vicsek physics up to a typical scale
$\xi_q$ that diverges exponentially with $q$, which we estimate using a
mean-field theory.  Finally, we use a scaling argument which does
  {\it not} depend on mean-field theory, but, rather, only on
  symmetry,
%However, this is only observed in large-enough systems: at fixed $q$
%one still observes the Vicsek scenario below a typical scale $\xi_q$
%that diverges exponentially with $q$ which we estimate using
%a mean-field theory. Finally, an energetic argument 
to trace back the fundamental difference with equilibrium 
%in the effect of anisotropy, compared to
%equilibrium system, 
to the presence of long-range order in the isotropic active system.

%%%%%%%%%%%%%%%%%%%%%%%%%%%%%%%%%%%%%%%
\begin{figure*}
  \centering
    \includegraphics[width=1.\linewidth]{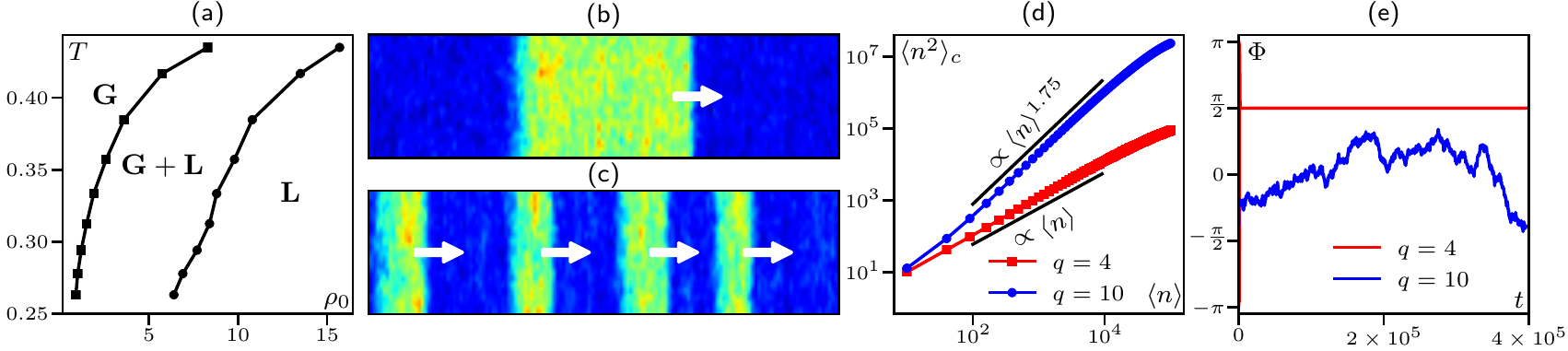}
  \caption{
    (a) Typical phase diagram in the $(\rho_0,T)$ plane ($q=4$), 
    Transition lines are defined by the coexisting densities at a given temperature $T=1/\beta$,
      computed in systems of size $400\times 20$).
(b,c): Snapshots of density field in the coexistence phase in a long $800 \times 20$ system suitable for the observation of many traveling bands (steady state, $\rho_0=10$, $q=4$ in (b), $q=10$ in (c)).   
(d) Number fluctuations $\langle n^2\rangle_{\rm c}$ vs $\langle n\rangle=\rho_0\ell^2$, the number of particles
 in a square box of linear size $\ell$ 
calculated in the liquid phase of square $200\times 200$ systems ($\rho_0=10$, $\beta=3.5$). 
``Giant'' anomalous fluctuations are observed for $q=10$, but not for $q=4$.
(e) Time series of $\Phi\equiv\arg \langle\vec m_k\rangle_k$, the orientation of the global polar order 
(parameters as in (d)).}
\label{FIG1}
\end{figure*}
%%%%%%%%%%%%%%%%%%%%%%%%%%%%%%%%%%%%%%%

{\it Active clock models.}  Particles $i=1,\ldots,N$ carrying a spin
$s_i\in\{0,1,\ldots,q-1\}$ reside at the nodes ${\bf R}$ of a square lattice
without occupation constraints.  They undergo biased diffusion by
jumping to neighboring sites with rate
$D(1+\eps \vec d\cdot \vec u_i)$ with $\vec d$ the direction of the
jump and $\vec u_i=(\cos\theta_i,\sin\theta_i)$ the unit vector along the clock angle
$\theta_i=2\pi s_i/q$.  Spins can rotate to the previous or next ``hour''
$\theta_{i}\pm\tfrac{2\pi}{q}$ at rate
\begin{equation}
    \label{eq:Wpm}
    w_{i,{\bf R}}=w_0 \exp\left[\tfrac{\beta}{2\rho_{\bf R}}\vec m_{\bf R}\cdot (\vec u_i'-\vec u_i)\right]
  \end{equation}
  where 
  %$S_{\bf k}$, 
  $\rho_{\bf R}$ and $\vec m_{\bf R}=\sum_{j\in{\bf R}} \vec u_j$ 
  are, respectively
  %, the set of spins,
  the number of particles and the magnetization at site ${\bf R}$
  hosting particle $i$, $\vec u_i'$ is the new spin direction, and
  $w_0$ is a constant \footnote{These rates are chosen such that if
    each site were isolated, the dynamics would satisfy detailed
    balance with steady-state probabilities
    $P_{\bf R}=\exp[-\beta H_{\bf R}]$ and
    $H_{\bf R}=-m_{\bf R}^2/(2\rho_{\bf R})$.  }.  For $q=2$, one
  recovers the AIM used in \cite{solon_flocking_2015}.  As shown
  in~\cite{SUPP}, in the isotropic $q\to\infty$ limit, the spin
  dynamics reduces to the Langevin equation
\begin{equation}
  \label{eq:langevin-theta}
  \partial_t \theta_i=\Omega_\infty + \sqrt{2D_\infty}\xi_i
\end{equation}
where $\xi_i$ is a Gaussian white noise of unit variance and the torque
%drift
and 
rotational diffusivity
%diffusion coefficients 
are given by
$\Omega_\infty=\frac{4w_0\pi^2\beta}{q^2}\left(\frac{\vec
    m_{\bf R}}{\rho_{\bf R}}\cdot\frac{\partial\vec
    u_i}{\partial\theta_i}\right)+O(q^{-3})$ and
$D_\infty=\frac{4w_0\pi^2}{q^2}+O(q^{-3})$, respectively. In order to have a
well-behaved active XY model in the $q\to\infty$ limit, one must thus
take $w_0\propto q^2$.  In the following, we set
$w_0=\tfrac{q^2}{4\pi^2}$ to fix $D_\infty=1$ and choose $D=1$
%, in both cases 
without loss of generality. For simplicity, we also fix the activity
parameter $\eps=0.9$. \footnote{For the AIM, it was shown that any amount of
activity $\eps>0$ leads to the same
phenomenology~\cite{solon_flocking_2015} and we have no reason to
suspect a different behavior in our active clock model.}

The only parameters left to vary, in addition to $q$, are thus the
temperature $T=1/\beta$ and the global density $\rho_0=N/(L_xL_y)$,
where $L_x$ and $L_y$ define a rectangular domain with periodic
boundary conditions. For numerical efficiency, we use parallel
updating, first performing on-site spin rotations, then biased
jumps.

Phase diagrams in the $(\rho_0,T)$ plane obtained for a given $q$
value all resemble those  
of the AIM or VM: the disordered gas present
at high $T$ and/or low $\rho_0$ is separated from the
low-$T$/high-$\rho_0$ polarly-ordered liquid by a coexistence phase
(Fig.~\ref{FIG1}(a)).  The liquid and coexistence phases both have a
finite global magnetization $m\equiv |\langle\vec m_{\bf R}\rangle_{\bf R}|$.
However, at fixed system size, they display AIM-like or VM-like
properties depending on $q$: For large-enough $q$, one observes giant
number fluctuations in the polar liquid and microphase separation as
for the Vicsek model (Fig.~\ref{FIG1}(c,d)). At lower $q$ values, on
the contrary, the liquid has normal fluctuations and the
system phase separates into a single moving domain
(Fig.~\ref{FIG1}(b,d)).  The global direction of order
$\Phi\equiv\arg \langle\vec m_{\bf R}\rangle_{\bf R}$ also behaves differently in
the liquid phase: it wanders slowly at high $q$, whereas it is pinned
at one of the clock angles at low $q$ (Fig.~\ref{FIG1}(e)).

%%%%%%%%%%%%%%%%%%%%%%%%%%%%%%%%%%%%%%%
\begin{figure}[b!]
  \centering
    \includegraphics[width=1\linewidth]{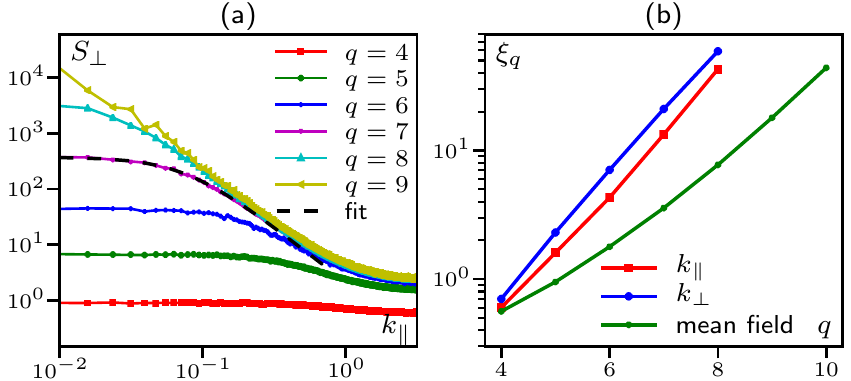}
  \caption{
(a) Liquid phase: $S_\perp(\vec q)=\langle m_\perp(\vec k)m_\perp(-\vec k)\rangle$ 
vs $\vec k=(k_\|, 0)$ for various $q$ values
($\rho_0=5.5$, $\beta=4$, system size $800\times 800$). 
For $q=7$ we show the function $\alpha/(1+(\xi k_\|)^2)$ with $\alpha$ and $\xi$ as fitted parameters. 
(b) Crossover length $\xi_q$ in the liquid phase extracted from 
(i) fits of $S_\perp(\vec k)$ as shown in (a) for $q=7$ for 
$\vec k=(k_\|, 0)$ and $\vec k=(0,k_\perp)$ (red and blue curves, respectively, same parameters as in (a)) 
and (ii) the mean-field prediction derived from Eq.~(\ref{eq:perturb-my}) (green curve).
}
\label{FIG2}
\end{figure}
%%%%%%%%%%%%%%%%%%%%%%%%%%%%%%%%%%%%%%%

%%%%%%%%%%%%%%%%%%%%%%%%%%%%%%%%%%%%%%% 
\begin{figure*}[t!] 
  \centering
  \includegraphics[width=1.01\linewidth]{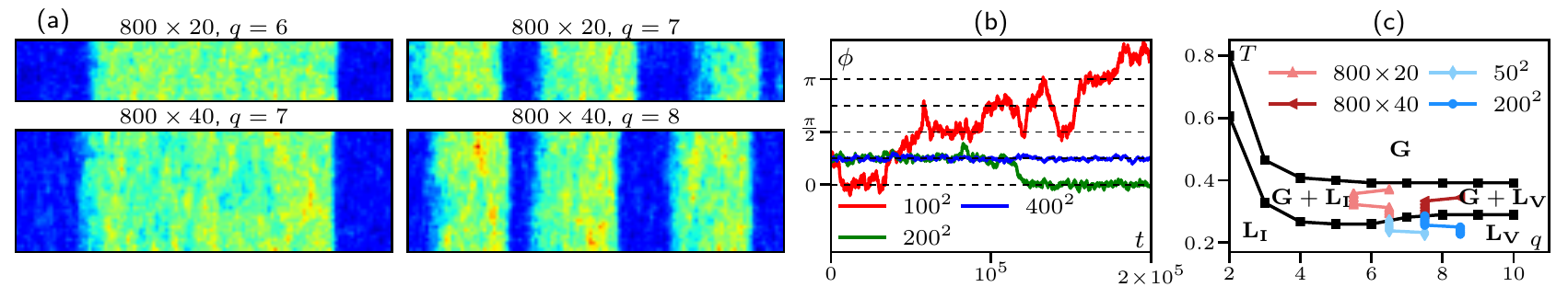}
  \caption{Transition from the active Ising to the Vicsek behavior as
    system size increases. {\bf (a)}: Snapshots of the density
    obtained after a long time $t=5\times 10^5$ starting from a large
    ordered band. The transition is shifted to larger $q$ as $L$
    increases. Same color code as in Fig.~\ref{FIG1}. $\beta=3.2$,
    $\rho_0=5.5$. {\bf (b)}: Direction of global order showing a
    transition between unpinned and pinned as system size
    increases. The dashed lines indicate the hours of the clock.
    $\beta=4$, $\rho_0=5.5$, $q=8$. {\bf (c)}: Phase diagram in the
    $q-T$ plane at $\rho_0=5.5$.  The line between macro- ($G+L_I$)
    and micro-phase separation ($G+L_V$) is defined as the transition
    between a single and multiple bands after time $t=10^6$ at the
    system size indicated in the legend.  The line separating the two
    liquids ($L_I$ and $L_V$) is defined as the transition between
    pinned and unpinned order parameter orientation after time
    $t=10^5$.}
  \label{FIG3}
\end{figure*}
%%%%%%%%%%%%%%%%%%%%%%%%%%%%%%%%%%%%%%% 

The results presented in Fig.~\ref{FIG1} seem to suggest that active clock
models have different behavior at $q=4$ and $q=10$, similar to the differences between
the AIM
and the VM.  In fact this is only true at finite size,
as perhaps best seen in the behavior of correlation functions in the
ordered liquid phase.  In Fig.~\ref{FIG2}(a), we show the transverse
magnetization structure factor
%$S_\rho(\vec k)=\langle \rho(\vec k)\rho(-\vec k)\rangle$ and
$S_\perp(\vec k)=\langle m_\perp(\vec k)m_\perp(-\vec k)\rangle$ for
wavelength $\vec k$ calculated in large systems for various $q$ values
(the same behavior is observed for the structure factor of the density
field). For sufficiently small $q$, $S_\perp$ converges to finite
values as $\vec k\to 0$.  This AIM-like behavior only happens, though,
beyond a crossover length scale $\xi_q$.  For scales smaller than
$\xi_q$, the structure factor exhibits algebraic scaling, as in the
VM.  The crossover scale $\xi_q$ can be extracted by fitting the
structure factors to the $\alpha/[1+(\xi_q k)^2]$ Ornstein-Zernicke
function~\footnote{While this fit cannot be perfect, since the
  structure factor of the isotropic model is
  expected~\cite{toner_long-range_1995,toner_flocks_1998,toner_hydrodynamics_2005,toner_reanalysis_2012}
  to scale like $k^{-\nu_{\parallel,\perp}}$ with neither
  $\nu_{\parallel}$ nor $\nu_\perp$ equal to $2$, both exponents are
  close enough to $2$ that this fit suffices to give a good estimate
  of $\xi_q$.}. This yields a typical scale $\xi_q$ that increases
exponentially rapidly with $q$ (Fig.~\ref{FIG2}(b)).  Extrapolating
these results, we expect that, even for large values of $q$, Vicsek
behavior will be observed up to (large) finite sizes, but that the
ultimate asymptotic behavior at the longest length scales is
Ising-like.

%Extrapolating these results, we can thus expect that even a very large
%system will only show Vicsek-like behavior in the liquid phase if $q$
%is sufficiently large, even though its actual asymptotic behavior is
%expected to be Ising-like.

Systems with linear size $L\ll \xi_q$ exhibit Vicsek phenomenology
with microphase separation and an unpinned order parameter in the
ordered phase. On the contrary, for $L\gg \xi_q$ the system shows
Ising behavior with full phase separation and pinned global
order. Fig.~\ref{FIG3}a illustrates this, showing that the transition
from microphase to macrophase separation happens upon increasing the
transverse system size at fixed $q$, whereas the reverse transition is
seen upon increasing $q$ at fixed system size.  Similarly, in the
liquid phase, there is a transition from unpinned to pinned order
parameter as $L$ increases (Fig.~\ref{FIG3}b).

The crossover from VM to AIM behavior can be summarized in the $(q,T)$
phase diagram at fixed global density (Fig.~\ref{FIG3}c). The three
expected phases are present, but one can, at a given system size,
define boundaries between Ising and Vicsek behavior within the
coexistence and the liquid phase regions, as described in the figure
caption.  These boundary lines
%, which are crossed when increasing $q$,
are displaced to higher and higher $q$ values as the system size is
increased.  Extrapolating to the infinite-size limit, VM-like behavior
is singular,
% or marginal?
confined to the infinite-$q$ (active XY) limit.

{\it Effective continuous description.} In equilibrium, clock models
are sometimes described at the field-theoretical level as continuous
spins subjected to an anisotropic external potential
$V_q(\phi)$~\cite{jose_renormalization_1977}, where $\phi$
parameterizes the local direction of order.  While usually postulated
on symmetry grounds, we have derived this potential at large $q$ using
a mean-field approximation~\cite{SUPP}, which yields
\begin{equation}
  \label{eq:potential-phi} V_q(\phi)=-\frac{2\rho}{\beta}\frac
  {I_q(\beta |\vec m|/\rho)}{ I_0(\beta |\vec m|/\rho)}\cos(q\phi)
\end{equation}
with $\vec m$ and $\rho$ the local magnetization and density,
$\phi=\arg(\vec m)$, and $I_n(x)$
% $I_n(x)=\sum_{k=0}^\infty(x/2)^{2k+n}\frac{1}{k!(k+n)!}$
the modified Bessel function of the first kind.
% This result
%, obtained using a mean-field approximation, 
%is expected to be a good
%approximation of the spin dynamics when the density $\rho$ is
%large. JT: no dynamics so far
The potential $V_q(\phi)$ is only the leading order 
contribution at large $q$, but a direct comparison with simulations of
the fully connected clock model shows that it is
already a good approximation for $q=4$~\cite{SUPP}.

We now demonstrate that we can understand the behavior of our
microscopic active clock model using the mean-field hydrodynamic
description of its isotropic, $q=\infty$ limit complemented by the anisotropic
potential~(\ref{eq:potential-phi}). This hydrodynamic theory,
derived in~\cite{SUPP} with standard techniques akin to those used for
the AIM~\cite{solon_flocking_2015}, yields
\begin{subequations}
  \begin{align}
    \partial_t \rho =& D\Delta\rho - v\nabla \cdot {\bf m} \label{eq:eq-rho}\\
    \partial_t {\bf m}   =& \left(\tfrac{\beta}{2}\!-\! 1 \!-\! \tfrac{\beta^2}{8\rho^2} {\bf m}^2\right) {\bf m}
+ D\Delta {\bf m} - \tfrac{\beta}{\rho} \,\partial_\phi V_q(\phi) \,{\bf m}^\perp\nonumber \\
 &+\tfrac{\beta v}{4\rho}( {\bf m}^\perp \nabla \cdot {\bf m}^\perp - {\bf m} \nabla \cdot {\bf m}) -\tfrac{v}{2}\nabla \rho\;,  \label{eq:eq-m}
  \end{align}
  \label{eq:hydro}
\end{subequations}
where ${\bf m}^\perp\equiv (-m_y,m_x)$.
\if{Using complex instead of vector
notation, i.e. $m=m_x+im_y$, $\triangledown=\partial_x +i\partial_y$,
and
$\vartriangle=\triangledown\triangledown^*=\partial_x^2+\partial_y^2$,
the resulting partial differential equations read:
\begin{subequations}
  \begin{align}
 \partial_t \rho =& D\!\vartriangle\!\!\rho - v\,{\rm Re}[ \triangledown^* m] \label{eq:eq-rho}\\
 \partial_t m   =& \left(\tfrac{\beta}{2}\!-\! 1 \!-\! \tfrac{\beta^2}{8\rho^2}|m|^2\right) m
+ D\!\vartriangle\!\!m-\tfrac{v}{2}\triangledown\rho \nonumber \\
 &-\tfrac{\beta v}{8\rho}\triangledown^* m^2 - i \tfrac{\beta}{\rho} \,\partial_\phi V_q(\phi) \,m  \label{eq:eq-m}
  \end{align}
  \label{eq:hydro}
    \end{subequations}
where $v=2D\varepsilon$ and the $\beta/\rho$ factor in front of
the potential is a mobility factor~\cite{SUPP}.}\fi

Consider now a perturbation of the homogeneous ordered state
%, assumed to be along the $x$ direction:
${\vec m}=(m_0+\delta m_\|,\delta m_\perp)$. To linear order, using
$\sin q\phi\approx q\phi\approx q m_\perp/m_\|$, we obtain for the $m_\perp$ field 
  \begin{equation}
    \label{eq:perturb-my} \dot{\delta m_\perp}=D\Delta \delta m_\perp
+\{\text{drift terms}\}-\alpha_q \delta m_\perp
\end{equation} with $\alpha_q=2q^2\tfrac {I_q(\beta m_0/\rho_0)}{
I_0(\beta m_0/\rho_0)}$. When $\alpha_q=0$, there is no mass on $m_\perp$,
as expected because of the continuous rotational symmetry. With
$\alpha_q>0$ however, a mass damps the fluctuations of $m_\perp$ and
therefore pins the direction of order. The typical length scale on
which this damping occurs is $\xi=\sqrt{D/\alpha_q}$. This scale
compares well to the crossover length $\xi_q$ measured in the microscopic
model (Fig.~\ref{FIG2}) albeit ---unsurprisingly--- not quantitatively.
%correct but it does show a similar exponential increase with $q$, and
%has the right order of magnitude.
  
%%%%%%%%%%%%%%%%%%%%%%%%%%%%%%%%%%%
\begin{figure}[b!]  \centering
\includegraphics[width=1\linewidth]{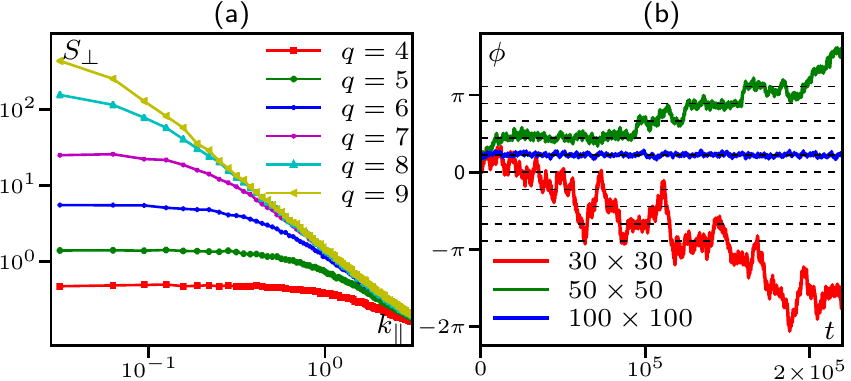}
    \caption{Ordered state in simulations of the PDE with noise
($\rho_0=5.5$, $\beta=4$, $D=1$, $v=1.8$) integrated with a space-time
discretization $dx=1$ and $dt=0.05$.  (a) Structure factor
$S_\perp(\vec k)=\langle m_\perp(\vec k)m_\perp(\vec -\vec k)\rangle$
vs $\vec k=(k_\|, 0)$ for various $q$ values (system size $200\times
200$).  (b) Time series of the orientation of global order $\Phi$
showing a transition between unpinned and pinned dynamics as system
size increases ($q=9$).}
    \label{FIG4}
\end{figure}
%%%%%%%%%%%%%%%%%%%%%%%%%%%%%%%%%%%

Let us now compare more precisely the behavior of
Eqs.~\eqref{eq:hydro} to our microscopic clock model.  To do that we
complement Eq.~(\ref{eq:eq-m}) with a white noise $\boldsymbol\eta(\vec r,t)$ of
random uniform orientation and Gaussian zero-mean unit-variance
amplitude,
% JT: It was wrong before since the equation was allegedly complex. It
% should be fine now, but a negative amplitude is not pretty. Is there
% a reason not to say "a centered Gaussian white
% noise with unit variance and independent components"?
and integrate the equations using a semi-spectral algorithm
(linear terms are computed in Fourier space, nonlinear ones in real
space) with Euler time stepping \footnote{Since we are interested in
the homogeneous ordered phase, we do not need to complement the mean
field equation with a density dependence of the coefficients, as
required to observe the bands of the coexistence
phase~\cite{solon_pattern_2015}.}.  The structure factor of $m_\perp$,
shown in Fig.~\ref{FIG4}(a), is found to be qualitatively similar to
that of Fig.~\ref{FIG2}. Moreover we also observe a pinning transition
when $L$ is increased (Fig.~\ref{FIG4}(b)), as in Fig.~\ref{FIG3} for
the microscopic models.

All in all, at sufficiently large sizes, anisotropy is always
relevant. This is markedly different from what happens in equilibrium
where, for $q>4$, there is a range of temperature over which the
anisotropy is irrelevant asymptotically, and one observes
quasi-long-range order, as for a continuous spin. The argument used to
derive a crossover length from the linearized hydrodynamic equation in
Eq.~(\ref{eq:perturb-my}) therefore fails in equilibrium. Indeed,
there is no homogeneous ordered state to perturb from, only a
quasi-ordered state with algebraically decaying correlations. This
difference is essential, as is clear from looking at the scaling with
system size of the energy $H_q=\int d^2\vec{r} \cos(q \theta(\vec r))$
due to the ``clock potential''.

Let us then compare the scaling of $\langle H_q\rangle_0$ with the
system size $L$, where the average is taken in the unperturbed system,
without $H_q$, in equilibrium and in the active case. Of course, we do
not actually {\it have} a Hamiltonian in our non-equilibrium model,
but this argument should roughly capture the effect of the potential
term in the equation of motion (\ref{eq:hydro}). Indeed, a more
rigorous dynamical RG calculation that makes no use of analogies with
equilibrium systems finds essentially the same result, as we will see
below.
In equilibrium, the unperturbed state 
can be described by the spin-wave Hamiltonian
$H_0=\int d^2\vec{r} \frac{K}{2}[\nabla \theta(\vec r)]^2$ with
stiffness $K$. The perturbation is then evaluated as
\begin{equation}
  \label{eq:Hq-eq}
  \langle H_q\rangle_0\approx \int d^2\vec{r}\langle e^{iq\theta(\vec r)}\rangle_0=\int  d^2\vec{r}e^{-\frac{q^2}{2} G(\vec 0)}
\end{equation}
where $G$ is the Green function of the Laplacian in infinite space.
In Fourier space $\hat G(\vec k)=\frac{T}{K k^2}$, which, in 2D,
gives $G(\vec 0)=T\log(L/\Lambda)/(2\pi K)$ for a system of size $L$, 
with $\Lambda$ a short distance
cut-off. Inserting this into Eq.~(\ref{eq:Hq-eq}), one obtains
\begin{equation}
  \label{eq:Hq-eq-scaling}
  \langle H_q\rangle_0\sim L^{2-\frac{q^2 T}{4\pi K}}.
\end{equation}
From Eq.~(\ref{eq:Hq-eq-scaling}), we see that anisotropy is relevant
when $L\to \infty$ whenever $T<T_q \equiv\frac{8\pi K}{q^2}$ and
irrelevant otherwise. As a result, if $T_1<T_{BKT}=\frac{\pi K}{4}$ which
happens for $q>q_c=4$, exactly, one observes a quasi-long range
ordered phase where the anisotropy is irrelevant for
$T_1<T<T_{\rm BKT}$, and a long-range ordered phase where the
anisotropy is relevant for $T<T_1$.

In the active case, the ordered state of the unperturbed system is
long-range ordered. Assuming that $\theta(\vec r)$ shows
Gaussian fluctuations with variance $\sigma^2$ around its mean value
$\theta_0$ leads to 
$\langle e^{iq\theta(\vec r)}\rangle_0=e^{-q^2\sigma ^2/2}$. In turn,
Eq.~(\ref{eq:Hq-eq}) becomes
\begin{equation}
  \label{eq:Hq-active}
  \langle H_q\rangle_0^{\rm active}\approx \int \vec{d^2r}\langle e^{iq\theta(\vec r)}\rangle_0=L^2 e^{-q^2\sigma^2/2}
\end{equation}
Anisotropy is thus always relevant when
$L\gg L_q\equiv e^{q^2\sigma^2/4}$, so that
$\langle H_q\rangle_0^{\rm active}\gg 1$. For $L\ll L_q$, on the
contrary, anisotropy is exponentially suppressed by $q$ and Vicsek
physics may be observed.

The argument above qualitatively explains the different responses
to anisotropy observed in the active and passive cases, which are
rooted in the different nature of the low temperature phases of their
isotropic limits. In equilibrium, the scaling argument presented above
has been made more rigorous using renormalization group
calculations~\cite{jose_renormalization_1977,elitzur_phase_1979}.  In
the active case, the essential conclusions can also be shown to hold,
following a dynamical renormalization group
analysis~\cite{toner_long-range_1995,toner_reanalysis_2012}.  This
analysis shows that the length scale $ L_{c\perp}$ beyond which the
symmetry breaking field changes the physics obeys~\cite{future}
\begin{equation}
  L_{c\perp}(q,\sigma)=a\exp\left({q^2\sigma^2\over 2z}\right) \,.
 \label{Lcperp}
\end{equation}
where $a$ is a microscopic length and $z$ a dynamic exponent whose
most recent numerical estimate is $z\simeq 1.33$~\cite{mahault_quantitative_2019}.

To summarize, we have shown that polar flocks are always susceptible
to spatial anisotropy at sufficiently large system sizes. This is
accompanied by a change of phenomenology compared to the isotropic
Vicsek case: the direction of the order parameter is pinned, not
wandering; structure factors at small $\vec k$ in the ordered phase
are constant, not diverging; and one has macrophase separation at
coexistence.  However, this is felt only beyond a characteristic
lengthscale that diverges for vanishingly small anisotropy. For the
$q$-state active clock model considered here, the crossover can be
understood from a hydrodynamic description where the discretization of
the spin direction is accounted for by an effective potential. The
crossover length is then shown to diverge exponentially with $q$, as
observed in the microscopic model. The difference with the passive
case can be accounted for using a scaling argument, that can be backed
up by renormalization group analysis, which shows that the presence of
long-range order is sufficient to render the anisotropy relevant
asymptotically for any value of $q$ and $T$.

Our study is a first step in understanding spatial anisotropy in
active matter. Its effect on other systems with a more complex
phenomenology, such as active rods or active nematics, will deserve
further investigations in the future. Finally, note that our choice of
a lattice model was made for numerical efficiency. We checked that our
results also hold in an off-lattice version of the model, but lack of
impact of the lattice, which is anisotropic in nature, is almost
surprising. How lattice anisotropy couples---or not---to the aligning
dynamics would surely deserve a further study, possibly revealing a
relevance at an even larger scale, out of reach of our numerical
simulations.

\textit{Acknowledgements:} We thank Mourtaza Kourbane-Houssene, for his early
involvement in this work, as well as Beno\^{\i}t Mahault for a
critical reading of the manuscript.

\bibliography{refs.bib}

\end{document}